\documentclass[12pt]{article}
\usepackage[dvips]{epsfig}
\usepackage[T1]{fontenc}
\usepackage[latin1]{inputenc}
\usepackage{graphicx}
\usepackage[english]{babel}
\usepackage{amsmath}
\usepackage{amssymb}
\usepackage{amsfonts}
\usepackage[T1]{fontenc}
\usepackage{color}
\usepackage{babel}
\usepackage{verbatim}
\usepackage[unicode=true,pdfusetitle,bookmarks=true,bookmarksnumbered=false,bookmarksopen=false,
 breaklinks=false,pdfborder={0 0 1},backref=false,colorlinks=true]{hyperref}
\hypersetup{linkcolor=blue,citecolor=blue}
\makeatletter
\usepackage[dvips]{epsfig}
\usepackage[T1]{fontenc}
\usepackage{color}
\usepackage{pdflscape}

\begin{document}

\title{\bf FLRW-Cosmology in Scalar-Vector-Tensor Theories of Gravity}

\author{Metin G{\" u}rses\thanks{%
email: gurses@fen.bilkent.edu.tr},~and~Yaghoub Heydarzade\thanks{%
email: yheydarzade@bilkent.edu.tr}
\\{\small Department of Mathematics, Faculty of Sciences, Bilkent University, 06800 Ankara, Türkiye}}

\date{\today}

\maketitle
\begin{abstract}
We generalize our previous theorem for FLRW spacetimes within the framework of generic metric gravity theories. In earlier work, we proved that, in the absence of matter fields, the field equations of any metric gravity theory constructed from the curvature tensor and its covariant derivatives reduce in FLRW spacetime to the Einstein equations with an effective perfect-fluid source.
In the present work, we extend this result to a broad class of scalar-vector-tensor theories in which the gravitational action contains arbitrary scalar and vector fields together with their covariant derivatives at any order. We prove that, under the symmetry conditions imposed by FLRW geometry, the metric field equations necessarily take the Einstein form with an effective perfect-fluid source, supplemented by the corresponding scalar and vector field equations.
This result shows that FLRW metrics belong to the class of universal metrics: the tensorial structure of the gravitational field equations is solely fixed by the symmetry of the FLRW spacetime and is independent of the specific form of the gravitation theory, while the resulting cosmological dynamics remains theory dependent. We illustrate our theorem using recently proposed Einstein-scalar and Einstein-Proca theories.
\end{abstract}

\section{Introduction}
Motivated by observational evidence, most notably the accelerated expansion of the universe, there has been growing interest in modifications of Einstein's theory of general relativity \cite{paddy}-\cite{odint}. One such extension is generic gravity theory, whose Lagrangian density incorporates all possible higher-curvature invariants as well as arbitrary higher-order covariant derivatives of the curvature tensor.
Certain classes of metrics, such as plane wave spacetimes, solve the field equations of generic gravity theories exactly \cite{Gibbons1}-\cite{Coley1}. These solutions are known as universal metrics. There also exist metrics that simplify the field equations without solving them identically. In particular, the Kerr-Schild-Kundt (KSK) class of metrics reduces the field equations of generic gravity theories to those of null dust with an effective cosmological constant \cite{Hervik1}-\cite{gst3}. Such spacetimes are referred to as almost universal metrics. Another important example of an almost universal spacetime is the Friedmann-Lemaitre-Robertson-Walker (FLRW) metric.

Recently, we showed that field equations of any generic gravity theory with FLRW metric reduce to the Einstein field equations with an effective perfect fluid energy-momentum tensor \cite{gur-hey-1, gur-hey-2}.
A generic metric gravity theory is a theory derivable from  the following
action using the variational principle
\begin{equation}\label{lag}
I=\int\, d^{D}\,x\, \sqrt{-g}\,\left (\frac{1}{\kappa} \left(R-2\Lambda \right)+
\mathcal{F}(g,\,\mbox{Riem},\,\nabla \mbox{Riem},\, \nabla \nabla \mbox{Riem}, \cdots)+ \mathcal{L}_{M} \right),
\end{equation}
where $g$, $\mbox{Riem}$, $\nabla \mbox{Riem}$, $\nabla \nabla \mbox{Riem}$, etc in $\mathcal{F}$ denote  the spacetime metric, Riemann tensor and its covariant derivatives at any order, respectively, and $\mathcal{L}_M$ is the Lagrangian of the matter
fields.  The function $\mathcal{F}(g,\,\mbox{Riem},\,\nabla \mbox{Riem},\, \nabla \nabla \mbox{Riem}, \cdots) $ is the part of the Lagrange function corresponding to higher order couplings, constructed from the metric, the Riemann tensor and its covariant derivatives. The corresponding field equations are
\begin{equation}
G_{\mu\nu}  +\Lambda g_{\mu\nu} + \mathcal{E}_{\mu \nu}= \kappa\,T_{\mu\nu}.
\end{equation}
Here,  $\mathcal{E}_{\mu \nu}$  is a symmetric divergent free tensor obtained from the variation of the scalar function $\mathcal{F}(g,\, \mbox{Riem},\,\nabla \mbox{Riem},\, \nabla \nabla \mbox{Riem}, \cdots)$ with respect to the spacetime metric $g_{\mu\nu}$. Our treatments in \cite{gur-hey-1, gur-hey-2} was to consider tensor $\mathcal{E}_{\mu \nu}$ as any symmetric second rank tensor obtained from the Riemann tensor and its covariant derivatives at any order. 

In the following,
we introduce the covariant description of the FLRW spacetimes in D-dimensions
and our mentioned theorem.
Let us begin with the definition of $D$-dimensional FLRW spacetime as follows.

\vspace{0.4cm}
\noindent
{\bf Definition}: {\it  The $D$-dimensional FLRW spacetime is defined with the following metric

\begin{equation}\label{metric}
g_{\mu\nu}=-u_\mu\, u_\nu+a^2\,h_{\mu\nu},
\end{equation}
where $\mu,\,\nu=0,...,D-1$, $a=a(t),~u_\mu=\delta^0_\mu$, and $h_{\mu\nu}$
reads as
\begin{equation}
h_{\mu\nu}=\begin{pmatrix}
0 & 0 & \hdots & 0 \\
0 &  &  &  \\
\vdots &  & h_{ij} &  \\
0 &  &  &  \\
\end{pmatrix},
\end{equation}
where $h_{ij}=h_{ij}(x^{a})$  with $i,j=1,...,D-1$ is the metric of a space of constant curvature $k$.}

\vspace{0.4cm}
\noindent
One can verify\begin{eqnarray}
&& u^\mu \,h_{\mu\nu}=u_\mu\, h^{\mu\nu}=0,\nonumber\\
&&h^{\mu}_{\alpha}= h^{\mu\alpha}\,h_{\alpha\nu}=\delta^{\mu}_{\nu}
+u^\mu \,u_\nu.
\end{eqnarray}
The corresponding Christoffel symbols to the metric (\ref{metric}) can be obtained as
\begin{equation}\label{christ}
\Gamma^\mu_{\alpha \beta}=\gamma^\mu_{\alpha\beta}-a\,\dot a\, u^\mu\, h_{\alpha \beta}+H\,\left( 2u_\alpha\, u^\mu\, u_\beta +u_\beta\, \delta^\mu_\alpha +u_\alpha\, \delta^\mu_\beta \right),
\end{equation}
where the dot sign represents the derivative with respect to time $t$, $H=\dot a/a$ is the Hubble parameter and $\gamma^\mu_{\alpha\beta}$ is defined as
\begin{equation}
\gamma^\mu_{\alpha\beta}=\frac{1}{2}a^2\, h^{\mu\gamma}\,\left(h_{\gamma \alpha,\beta}+h_{\gamma \beta,\alpha} -h_{\alpha \beta,\gamma} \right).
\end{equation}
One can also prove the following properties for $u_\alpha$ and $h_{\alpha\beta}$
\begin{eqnarray}
u_\mu\,h^\mu_{\alpha\gamma,\beta}&=&0= u_\mu\,\gamma^\mu_{\alpha\beta},\nonumber\\
\nabla_{\alpha}u_\beta&=&-a\,\dot a \,h_{\alpha\beta}=-H\left(g_{\alpha \beta}+u_\alpha\,
u_\beta  \right),\label{u}\nonumber\\
\nabla_{\gamma}h_{\alpha\beta}&=&-H\left(2u_\gamma\, h_{\alpha\beta}+u_\beta\, h_{\gamma\alpha}  +u_\alpha\, h_{\gamma\beta}    \right)\nonumber\\
&=&-\frac{\dot a}{a^3}\left(2u_\gamma \,g_{\alpha\beta}+u_\beta\, g_{\gamma\alpha}  +u_\alpha\, g_{\gamma\beta}  +4u_\alpha\, u_\beta\, u_\gamma \right).
\end{eqnarray}
These construct a closed tensor algebra for the FLRW geometry. We find that  the  Riemann curvature tensor and its covariant derivative can be written  in the following
linear form in terms of the metric  $g_{\mu\nu}$ and the four vector
$u_\mu$
\begin{eqnarray}\label{riemann2}
R_{\mu \alpha\beta\gamma}&=&\left(  g_{\mu_\beta} g_{\alpha\gamma}-g_{\mu_\gamma} g_{\alpha\beta}\right)\rho_1+\left(u_{\mu}\left(g_{\alpha\gamma} u_\beta -  g_{\alpha\beta} u_\gamma \right)-u_\alpha\,\left(g_{\mu_\gamma} u_\beta -  g_{\mu_\beta} u_\gamma \right)\right)\rho_2,\nonumber\\
\nabla_\lambda R_{\mu\alpha\beta\gamma}
&=&
\dot{\rho}_1\,u_\lambda
\left(g_{\mu\beta}g_{\alpha\gamma}-g_{\mu\gamma}g_{\alpha\beta}\right)
\nonumber\\
&&+
\dot{\rho}_2\,u_\lambda
\left[
u_\mu\left(g_{\alpha\gamma}u_\beta-g_{\alpha\beta}u_\gamma\right)
-
u_\alpha\left(g_{\mu\gamma}u_\beta-g_{\mu\beta}u_\gamma\right)
\right]
\nonumber\\
&&
-H\rho_2\Big[
g_{\lambda\mu}\left(g_{\alpha\gamma}u_\beta-g_{\alpha\beta}u_\gamma\right)
+u_\mu\left(g_{\alpha\gamma}g_{\lambda\beta}-g_{\alpha\beta}g_{\lambda\gamma}\right)
\nonumber\\
&&\qquad
-g_{\lambda\alpha}\left(g_{\mu\gamma}u_\beta-g_{\mu\beta}u_\gamma\right)
-u_\alpha\left(g_{\mu\gamma}g_{\lambda\beta}-g_{\mu\beta}g_{\lambda\gamma}\right)
\nonumber\\
&&\qquad
+2u_\lambda
\left[
u_\mu\left(g_{\alpha\gamma}u_\beta-g_{\alpha\beta}u_\gamma\right)
-
u_\alpha\left(g_{\mu\gamma}u_\beta-g_{\mu\beta}u_\gamma\right)
\right]
\Big],
\end{eqnarray}
where $\rho_1$ and $\rho_2$ are defined as
\begin{eqnarray}\label{rho1}
&&\rho_1=H^2 +\frac{k}{a^2},\nonumber\\
&&\rho_2=H^2 +\frac{k}{a^2}-\frac{\ddot a}{a}=-\dot{H}+\frac{k}{a^2}.
\end{eqnarray}
Hence, any geometric tensor (at any order of covariant derivatives and products
of Riemann tensor) in FLRW geometry can be expressed as the  sum of monomials
composed of product of $u_\mu$ and $g_{\mu\nu}$.

The contractions of the Riemann tensor (\ref{riemann2}) gives the Ricci tensor
and Ricci scalar, respectively, as
\begin{eqnarray}\label{ricci}
&&R_{\alpha\gamma}=g_{\alpha\gamma}\left( (D-1)\rho_1-\rho_2 \right)+u_\alpha u_\gamma(D-2)\rho_2, \nonumber\\
&&R=\left(D-1 \right)\left(D\rho_1 -2\rho_2 \right).
\end{eqnarray}
Using the expressions in \eqref{riemann2}-\eqref{ricci}, one can easily
verify that  FLRW spacetimes are locally conformally flat for all values of spatial curvature $k$ in any dimensions, i.e. the Weyl tensor vanishes identically \cite{chen, mant1, mant2, mant3}.

Hence, we have the following theorem.

\vspace{0.3cm}
\noindent
{\bf Theorem 1}: {\it Any second rank tensor obtained from the metric tensor, the Riemann tensor,  Ricci tensor, and their covariant derivatives at any order is a combination of the metric tensor $g_{\mu \nu}$ and $u_{\mu} u_{\nu}$ that is
\begin{equation}\label{E}
\mathcal{E}_{\mu \nu}=A g_{\mu \nu}+B u_{\mu} u_{\nu},
\end{equation}
where $A$ and $B$ are functions of $a(t)$ and their time derivatives at any order.}

\vspace{0.5cm}
\noindent
Some special cases of this theorem are given in \cite{capoz1, capoz2, capoz3}. In these references, this theorem was proved for the special cases $\mathcal{F}(R,\mathcal{G})$ and $\mathcal{F}(R,\, \square R,\,\square \square R, \cdots)$ field equations. In \cite{capoz3} the geometry  is the generalized FLRW spacetime.

We have the following corollary of the above theorem.
\vspace{0.3cm}
\noindent
\\
\textbf{Corollary 1: } The field equations of any generic gravity theory
take the form
\begin{equation}\label{FEs}
G_{\mu \nu}+ \Lambda g_{\mu \nu}+\mathcal{E}_{\mu \nu}=\kappa\,T_{\mu \nu},
\end{equation}
where $G_{\mu \nu}$ is  the Einstein tensor, $\Lambda$ is the cosmological constant, $T_{\mu \nu}$ is the ordinary energy-momentum tensor of perfect fluid distribution and $\mathcal{E}_{\mu \nu}$ comes from the higher order curvature terms.   Hence, the general field
equations read as
\begin{eqnarray}\label{rp}
&& \kappa\,\rho=\frac{1}{2}(D-1)(D-2) \rho_{1}-\Lambda +B-A, \label{gen1}\\
&& \kappa\,p=(D-2)\left[-\frac{1}{2}(D-1) \rho_{1}+\rho_{2}\right] +\Lambda+A. \label{gen2}
\end{eqnarray}

The present work  aims to extend the above theorem when extra
matter fields exist. In particular, it may be highly interesting when  scalar and vector fields are included in the generic geometric Lagrangian (\ref{lag}). However, we
would like to bring the following remark to the attention of the reader.
\\
\\
\textbf{Remark:} The main claim of the present study is not  to only extend the above theorem when extra matter fields exist. The result here is conceptually different and much stronger. We do not start from a particular theory and perform a model-dependent rearrangement of its cosmological equations. Instead, we prove a general theorem stating that for a very broad class of gravitational theories the metric field equations on an FLRW background necessarily reduce to the Einstein equations with an effective perfect-fluid source.
This result is closely related to the concept of \emph{universal metrics}
and turns out that FLRW metric belongs to the class of universal metrics.

For this purpose, in the next section, we generalize the Theorem 1 when we consider scalar-tensor and vector-tensor theories, and show that the equations (\ref{gen1}) and (\ref{gen2}) remains the same but the functions $A$ and $B$ may contain scalar and  vector fields and their covariant derivatives. In the subsections 2.1 and 2.2, we present examples of scalar-tensor and vector-tensor theories, respectively.

\section{Extension of the FLRW Tensor Algebra with Scalar and Vector Fields}

In our previous work \cite{gur-hey-1} outlined in the previous section , we established that, for the $D$-dimensional FLRW spacetime,
the curvature sector admits a closed tensor algebra. In particular, in Theorem~1 we
 showed that any symmetric rank-2 tensor constructed from the metric,
curvature tensors, and their covariant derivatives at any order can be written as a
linear combination of the metric tensor $g_{\mu\nu}$ and the multiple of timelike unit vectors
$u_\mu u_\nu$, with coefficients depending only on the scale factor $a(t)$ and its
derivatives. This led to the universal perfect-fluid structure of the effective field
equations in any higher-order metric gravity theories.

\vspace{0.5cm}
\noindent
We now aim to extend these results to the scalar-vector theories including the contributions of a homogeneous scalar field
$\phi=\phi(t)$ and an isotropic vector field ansatz $A_\mu=\psi(t)u_\mu$, together with
their covariant derivatives at any order and possible non-minimal couplings to the curvature.
Our aim is to prove that the closure of FLRW algebra and the reduction to the $\{g_{\mu\nu},\,u_\mu u_\nu\}$ basis persists
in this more general setting.

\vspace{0.5cm}
\noindent
\textbf{Sketch of the Proof:} Let us begin with the scalar sector closure.
For a homogeneous scalar $\phi(t)$ one readily computes
\begin{equation}
\nabla_\mu \phi = \dot\phi\,u_\mu, \qquad
\nabla_\mu\nabla_\nu \phi = \ddot\phi\,u_\mu u_\nu + H\dot\phi\,(g_{\mu\nu}+u_\mu u_\nu).
\end{equation}
where $H=\dot a(t)/ a(t)$. Any higher covariant derivative of $\phi$ is again expressible as a linear combination of
monomials built from $g_{\mu\nu}$ and $u_\mu=\delta^0_\mu$, with time-dependent coefficients depending
on scale factor $a(t)$ and $\phi(t)$ and their time derivatives. Thus, the scalar sector obeys the same closure rules
as the curvature tensors.

For the vector sector closure, we consider the isotropic ansatz $A_\mu=\psi(t)u_\mu$. Using the relations
\[\nabla_\mu u_\nu=H(g_{\mu\nu}+u_\mu u_\nu),~~~~\nabla_\mu u^\mu=(D-1)H, \]
one finds
\begin{equation}
\nabla_\mu A_\nu = \dot\psi\,u_\mu u_\nu + H\psi\,(g_{\mu\nu}+u_\mu u_\nu),
~~~~\nabla_\mu A^\mu = -\dot\psi + (D-1)H\psi.
\end{equation}
Hence,
\begin{equation}
\nabla_{(\mu}A_{\nu)} = \alpha(t)g_{\mu\nu}+\beta(t)u_\mu u_\nu,
\qquad \nabla_{[\mu}A_{\nu]}=0,
\end{equation}
for the time dependent functions $\alpha$ and $\beta$. Therefore the symmetrized derivative produces
no new structures beyond $g_{\mu\nu}$ and $u_\mu u_\nu$, while the antisymmetric
derivative vanishes identically for the timelike ansatz. Higher derivatives
$\nabla\cdots\nabla A$ preserve this pattern, leading to no additional
independent tensors.

Combining the curvature, scalar, and vector sectors, we conclude that all building
blocks --- $g_{\mu\nu}$, curvature tensors, $\phi$, $A_\mu$, and their covariant
derivatives at any order --- reduce to finite sums of monomials of rank $k$ as

\begin{equation}
M_{\mu_{1} \mu_{2} \mu_{3} \mu_{4} \cdots  \mu_{k}}=g_{\mu_{1} \mu_{2}} g_{\mu_{3} \mu_{4}} \cdots u_{\mu_{k-1}} u_{\mu_{k}},
\end{equation}
with time-dependent coefficients functions. Here, there are $r$ number of metric tensor and $k-2r$ number of vector $u_{\mu}$ in a monomial of rank $k$ and  $r$ is any nonnegative integer.
Hence, if  $E_{\alpha_{1} \alpha_{2} \cdots \alpha_{m}}$ is a tensor of
rank $m$ obtained from the curvature tensors, scalar and vector fields, and their covariant derivatives at any order, then,  it takes the following form for $m= \mbox{even integer}$
\begin{eqnarray}
E_{\alpha_{1} \alpha_{2} \cdots \alpha_{m}}&=& A_{1}\, g_{\alpha_{1} \alpha_{2}} \cdots g_{\alpha_{m-1} \alpha_{m}} +A_{2}\, g_{\alpha_{1} \alpha_{2}} \cdots u_{\alpha_{m-1}} u_{\alpha_{m}}+\cdots  \nonumber\\
&&+A_{m-1}\, g_{\alpha_{1} \alpha_{2}}\, u_{\alpha_{3}} \cdots u_{\alpha_m}+A_{m}\, u_{\alpha_{1}}\, u_{\alpha_{2}} \cdots u_{\alpha_{m}},
\end{eqnarray}
and for $m=\mbox{odd integer}$ as
\begin{eqnarray}
E_{\alpha_{1} \alpha_{2} \cdots \alpha_{m}}&=& B_{1}\, g_{\alpha_{1} \alpha_{2}} \cdots g_{\alpha_{m-2} \alpha_{m-1}}\,u_{\alpha_{m}} +B_{2}\, g_{\alpha_{1} \alpha_{2}} \cdots u_{\alpha_{m-2}}\, u_{\alpha_{m-1}} u_{\alpha_{m}}+\cdots  \nonumber\\
&&+B_{m-1}\, g_{\alpha_{1} \alpha_{2}}\, u_{\alpha_{3}} \cdots u_{\alpha_m}+B_{m}\, u_{\alpha_{1}}\, u_{\alpha_{2}} \cdots u_{\alpha_{m}},
\end{eqnarray}
where $A_{k},B_{k}$ ($k=1,2, \cdots , m$) are functions of the time parameter
$t$. All the tensors of rank two obtained by the contraction of such tensors are of our interests.  Since $u_{\alpha}\, u^{\alpha}=-1$ and $g_{\mu\nu}$ is the metric tensor then any second rank tensor obtained from the contraction of such two different monomials is either
$g_{\mu \nu}$ or   $u_{\mu} u_{\nu}$. Therefore, if $E_{\mu \alpha_{1} \alpha_{2} \cdots \alpha_{m}}$ and $F_{\nu}\,^{\alpha_{1} \alpha_{2} \cdots \alpha_{m}}$ are two tensors obtained from the curvature tensors, scalar and vector fields,
and their covariant derivatives at any order, then
we have
\begin{equation}
 E_{\mu \alpha_{1} \alpha_{2} \cdots \alpha_{m}}\,F_{\nu}\,^{\alpha_{1} \alpha_{2} \cdots \alpha_{m}}=C_{1}\, g_{\mu \nu}+C_{2} u_{\mu} u_{\nu},\\  \label{ozel}
\end{equation}
where $C_{1}$ and $C_{2}$ are some scalars in time parameter $t$.
This means that the space of rank-2 tensors produced by index contractions of
FLRW-invariant monomials is the span of $\{\, g_{\mu\nu},\; u_\mu u_\nu \,\}$.
This statement also covers mixed
monomials and any number of covariant time derivatives. Hence, we have the following theorem and its corollary.
\\
\\
{\bf Theorem 2:}
On the $D$-dimensional FLRW background, with homogeneous scalar field $\phi=\phi(t)$
and isotropic vector ansatz $A_\mu=\psi(t)u_\mu$, any symmetric rank-2 tensor constructed
from $g_{\mu\nu}$, the curvature tensors, $\phi$, $A_\mu$, and any number of their
covariant derivatives, including their arbitrary non-minimal couplings, can be written as
\begin{equation}
\mathcal{X}_{\mu\nu}=A(t)\,g_{\mu\nu}+B(t)\,u_\mu u_\nu,
\end{equation}
where $A$ and $B$ are time-dependent functions built from
$a(t),\phi(t), \psi(t)$ and their derivatives.
\\
\\
{\bf Corollary 2:}
Consider a generic non-minimally  coupled scalar-vector-tensor gravity theory by the action
\begin{equation}
I=\frac{1}{2\kappa}\int\, d^{D}\,x\, \sqrt{-g}\,\left (\mathcal{L}_1+\mathcal{L}_2+\mathcal{L}_3+\mathcal{L}_4\right)
+\int\, d^{D}\,x\, \sqrt{-g}\, \mathcal{L}_{M} ,
\end{equation}
where $\kappa=8\pi G c^{-4}$ is the gravitational coupling\footnote{We will
work in the units $\kappa =1$. }, and the total Lagrange function includes \begin{eqnarray}
\mathcal{L}_1&=& \mathcal{L}_1(g_{\mu\nu}, \mbox{Riem}, \nabla \mbox{Riem}, \nabla \nabla \mbox{Riem},...),\\
\mathcal{L}_2&=& \mathcal{L}_2(\phi, \nabla \phi, \nabla \nabla \phi,...),\\
\mathcal{L}_3&=& \mathcal{L}_3(A, \nabla A, \nabla\nabla A,...),\\
\mathcal{L}_4&=& \mathcal{L}_4(g, \mbox{Riem}, \phi, A, \nabla \mbox{Riem}, \nabla \phi,
\nabla A, ...),
\end{eqnarray}
in which $\mathcal{L}_1, \mathcal{L}_2, \mathcal{L}_3, \mathcal{L}_4$ and $\mathcal{L}_M$
are the geometric, scalar field, vector field, any non-minimal coupling among
these, and matter Lagrangian
functions,
respectively,  in terms of the metric $g_{\mu\nu}$, scalar field $\phi$, and vector field $A^\mu$.
  The corresponding field equations by the variation with respect to the metric, scalar and vector
fields
 can be written as
\begin{eqnarray}
G_{\mu\nu} + \mathcal{E}_{\mu \nu}&=&T_{\mu\nu},\label{fess}\\
f_1\left(\phi,...,\phi^{(n)},\mbox{curvature scalars},\psi,...,\psi^{(m)} \right)&=&0,\\
f_2\left(\psi,...,\psi^{(l)},\mbox{curvature scalars},\phi,...,\phi^{(k)} \right)&=&0,
\end{eqnarray}
where   $\mathcal{E}_{\mu \nu}$  denotes all higher-curvature, scalar, and vector contributions
from $\mathcal{L}_{i}, i=1,2,3,4$, and $f_1$ and $f_2$ are two functions representing the equations of motion of non-minimally coupled curvature, scalar and vector fields with $m, n, l, k$ representing the derivative order. By the Theorem 1, $\mathcal{E}_{\mu\nu}$ must take the form
\begin{equation}
\mathcal{E}_{\mu\nu}=A(t)\,g_{\mu\nu}+B(t)\,u_\mu u_\nu.
\end{equation}
Hence,  the field equations \eqref{fess} will be
\begin{align}
\rho &= \tfrac{1}{2}(D-1)(D-2)\rho_1+B-A, \\
p &= (D-2)\left[-\tfrac{1}{2}(D-1)\rho_1+\rho_2\right]+A,
\end{align}
where $\rho$ and $p$ are the energy density and pressure of the ordinary matter
energy-momentum tensor $T_{\mu\nu}=(\rho+p)u_\mu u_\nu +pg_{\mu\nu}$, and the scalar function $\rho_{1}$ and $\rho_{2}$ are defined as in \eqref{rho1}.
Thus $A$ behaves as a pressure term, while $B-A$ contributes like as a density. This interpretation generalizes the
earlier geometric-only result to include the new scalar field and vector field sectors \cite{gur-hey-2}.

\vspace{0.5cm}
\noindent
 The novelty of the present work is not the introduction of an effective-fluid description for a particular modified-gravity model. Rather, we prove that this perfect-fluid structure is \emph{inevitable} for an extremely broad class of scalar-vector-tensor theories once FLRW symmetry is imposed. In this sense, our result is structural and model-independent, rather than a theory-by-theory reformulation.

\vspace{0.5cm}
\noindent
We also emphasize that many well-known examples in the literature, such as $f(R)$ gravity, scalar-tensor models, and Gauss-Bonnet theories, appear naturally as special cases within this general framework. The examples studied in the present manuscript therefore serve as illustrations of the general theorem rather than independent derivations of the effective-fluid form.

\vspace{0.5cm}
\noindent
In the following, we shall give two specific examples for the application of the
 theorem and its corollary in scalar-tensor and vector-tensor theories.
 
 \subsection{Regularized  Scalar-Tensor 4D Einstein-Gauss-Bonnet  Theory}
Here, we consider the regularized four-dimensional scalar-tensor Einstein--Gauss--Bonnet (EGB) theory introduced in \cite{Fernandes2020}. The $D$-dimensional EGB action with matter is given by
\begin{equation}
S_D[g_{\mu\nu}] \;=\; \int d^D x \sqrt{-g}\,\bigl(R + \alpha\,\mathcal{G}\bigr) + S_m \, ,
\label{eq:EGB_D_dim_intro}
\end{equation}
where $\mathcal{G} = R_{\mu\nu\rho\sigma} R^{\mu\nu\rho\sigma} - 4 R_{\mu\nu} R^{\mu\nu} + R^2$ is the Gauss--Bonnet invariant and $\alpha$ is a coupling constant and $S_m$ is the matter action. In $D=4$, the Gauss--Bonnet term is topological and does not contribute to the local field equations, so a non-trivial $4D$ limit is sought by introducing a dimensional-regularization-like rescaling of the coupling,
\begin{equation}
\alpha \;\to\; \frac{\hat{\alpha}}{D-4} \, ,
\label{eq:alpha_rescale_intro}
\end{equation}
and then considering the limit $D \to 4$ \cite{Glevan}. In \cite{gurses-sisman-tekin,
gurses-tekin},
it was shown that the field equations are not defined as $D\to 4$. Hence,
this prescription renders the action ill-defined, and a regularization procedure is therefore required. Following the two-dimensional construction in \cite{Mann1993}, the authors of \cite{Fernandes2020} introduced a conformally related metric
\begin{equation}
\tilde{g}_{\mu\nu} = e^{2\phi} g_{\mu\nu} \, ,
\label{eq:conformal_metric_intro}
\end{equation}
with scalar field $\phi$, and added to \eqref{eq:EGB_D_dim_intro} the following
counter
term
\begin{equation}
S_{\mathrm{ct}} = - \alpha \int d^D x \sqrt{-\tilde{g}}\,\tilde{\mathcal{G}} \, ,
\label{eq:counterterm_intro}
\end{equation}
where $\tilde{\mathcal{G}}$ is the Gauss--Bonnet invariant constructed from $\tilde{g}_{\mu\nu}$. Using the conformal transformation of the Gauss--Bonnet term in $D$ dimensions, one can write
\begin{equation}
\sqrt{-\tilde{g}}\,\tilde{\mathcal{G}}
= \sqrt{-g}\,\Bigl[\mathcal{G} + (D-4)\,\mathcal{X}[g,\phi]+\nabla_\mu J^\mu[g,\phi] + \mathcal{O}\bigl((D-4)^2\bigr)\Bigr] \, ,
\label{eq:GB_expansion_intro}
\end{equation}
for some scalar density $\mathcal{X}[g,\phi]$ and vector $J^\mu [g,\phi]$ built from $g_{\mu\nu}$, $\phi$ and their derivatives. Substituting \eqref{eq:GB_expansion_intro} into \eqref{eq:counterterm_intro} and combining with \eqref{eq:EGB_D_dim_intro}, the total Gauss--Bonnet sector of the action becomes
\begin{eqnarray}
\int d^D x \sqrt{-g}\,\alpha\,\mathcal{G} \;+\; S_{\mathrm{ct}}
&=&
\int d^D x\,\alpha\,\left( \sqrt{-g}\mathcal{G}- \sqrt{-\tilde{g}}\,\tilde{\mathcal{G}} \right) \nonumber\\
&=&- \int d^D x \sqrt{-g}\alpha \left( (D-4)\mathcal{X}[g,\phi] +\nabla_\mu J^\mu [g,\phi]+ \mathcal{O}\left( (D-4)^2 \right) \right).
\end{eqnarray}
After inserting the rescaling \eqref{eq:alpha_rescale_intro}, the overall factor of $(D-4)$ cancels, and the limit $D \to 4$ exists and is finite. In the four-dimensional limit, an explicit computation using the known conformal transformation properties of the curvature tensors and integrations by parts then gives
\begin{equation}
\mathcal{X}[g,\phi] = - \Bigl( 4 G^{\mu\nu}\nabla_\mu\phi\nabla_\nu\phi - \phi\,\mathcal{G}
+ 4 \Box\phi\,(\nabla\phi)^2 + 2 (\nabla\phi)^4 \Bigr) \, ,
\end{equation}
so that the regularized $4D$ theory is described by the action
\begin{equation}
S_{\mathrm{reg}}[g_{\mu\nu},\phi] = \int d^4 x \sqrt{-g}\,
\Bigl[ R + \hat{\alpha}\,\bigl(4 G^{\mu\nu}\nabla_\mu\phi \nabla_\nu\phi
- \phi\,\mathcal{G}
+ 4 \Box\phi\,(\nabla\phi)^2
+ 2 (\nabla\phi)^4 \bigr) \Bigr] + S_m \, .
\label{eq:regularized_action_4D_intro}
\end{equation}

The field equations of the regularized theory follow from varying \eqref{eq:regularized_action_4D_intro} with respect to $g_{\mu\nu}$ and $\phi$. Variation with respect to the metric yields
 \begin{equation}\label{fescalar}
 G_{\mu\nu} = \hat{\alpha} \, \hat{H}_{\mu\nu} + T_{\mu\nu} ,
\end{equation}
where  $T_{\mu\nu}$ is the usual matter stress--energy tensor from $S_m$ and $\hat{H}_{\mu\nu}$ is a symmetric tensor containing at most second derivatives of $g_{\mu\nu}$ and $\phi$ given as
\begin{eqnarray}
\hat{H}_{\mu\nu} &=&
2 R \left( \nabla_\mu \nabla_\nu \phi - \nabla_\mu \phi \nabla_\nu \phi \right)
+ 2 G_{\mu\nu} \left( (\nabla \phi)^2 - 2 \Box \phi \right)
+ 4 G_{\nu}{}^{\alpha} \left( \nabla_\alpha \nabla_\mu \phi - \nabla_\alpha \phi \nabla_\mu \phi \right)\nonumber\\
&&+ 4 G_{\mu}{}^{\alpha} \left( \nabla_\alpha \nabla_\nu \phi - \nabla_\alpha \phi \nabla_\nu \phi \right)
+ 4 R_{\mu\alpha\nu\beta}
    \left( \nabla^\beta \nabla^\alpha \phi - \nabla^\alpha \phi \nabla^\beta \phi \right)\nonumber\\
&&+ 4 \nabla_\alpha \nabla_\nu \phi
    \left( \nabla^\alpha \phi \nabla_\mu \phi - \nabla^\alpha \nabla_\mu \phi \right)+ 4 \nabla_\alpha \nabla_\mu \phi \, \nabla^\alpha \phi \nabla_\nu \phi\nonumber\\
&&- 4 \nabla_\mu \phi \nabla_\nu \phi
    \left( (\nabla\phi)^2 + \Box\phi \right)
+ 4 \Box\phi \, \nabla_\nu \nabla_\mu \phi
\nonumber\\
&&- g_{\mu\nu} \Big[
    2 R (\Box\phi - (\nabla\phi)^2)
 + 4 G_{\alpha\beta}
        \left( \nabla^\beta \nabla^\alpha \phi
             - \nabla^\alpha \phi \nabla^\beta \phi \right)
    \nonumber\\
    &&~~~~~~~~+ 2 (\Box\phi)^2
    - (\nabla\phi)^4
+ 2 \nabla_\beta \nabla_\alpha \phi
        \left( 2 \nabla^\alpha \phi \nabla^\beta \phi
             - \nabla^\beta \nabla^\alpha \phi \right)
  \Big].
\end{eqnarray}
Variation of the action with respect the scalar field $\phi$ gives
\begin{equation}\label{scalar}
R_{\mu\nu} \nabla^\mu \phi \nabla^\nu \phi
- G_{\mu\nu} \nabla^\mu \nabla^\nu \phi
- \Box\phi \, (\nabla\phi)^2
+ (\nabla_\mu \nabla_\nu \phi)(\nabla^\mu \nabla^\nu \phi)
- (\Box\phi)^2
- 2 \nabla_\mu \phi \nabla_\nu \phi
    \nabla^\mu \nabla^\nu \phi
= \frac{1}{8} \, \mathcal{G}.
\end{equation}

In order to apply our theorem to the field equations of 4D regularized EGB theory, remember that for the FLRW spacetime, the Riemann tensor,   Ricci tensor and
Ricci scalar, are expressed solely by the metric tensor  $g_{\mu\nu}$ and the timelike vector $u_{\mu}$ as in \eqref{riemann2} and \eqref{ricci}.
\\
Hence, using
\begin{eqnarray}
\nabla_\mu \phi&=& \dot \phi u_\mu,~~~~~~\nabla_\mu \nabla_\nu \phi =\ddot \phi +H\dot \phi (g_{\mu\nu}+u_\mu u_\nu),\nonumber\\
 \Box \phi&=&-\ddot \phi +3H\dot \phi,~~~~~~(\nabla\phi)^2=-\dot\phi^2,
\end{eqnarray}
we find $\hat{H}_{\mu\nu}$ tensor in the form
 \begin{equation}
\hat{H}_{\mu\nu} = A(t)g_{\mu\nu}
+ B(t)u_\mu u_\nu,
\end{equation}
where the explicit form of the scalar functions $A(t)$ and $B(t)$ are
\begin{align}
A(t) &=
\dot{\phi}^{4}
- 4\,\dot{\phi}^{2}\ddot{\phi}
+ 8H\dot{\phi}\ddot{\phi}
- 4H^{2}\ddot{\phi}
+ 8H^{3}\dot{\phi}
- 6H^{2}\dot{\phi}^{2}
- \dot{H}\left(4\dot\phi^2-8H\dot \phi\right)+\frac{k}{a^2}\left(2\dot\phi^2
-4\ddot \phi \right),\nonumber\\
&=\dot\phi^{\,4}
+\bigl(8H\dot\phi-4\dot\phi^{\,2}-4H^2-4\frac{k}{a^2}\bigr)\ddot\phi
+\bigl(8H^3+8H\dot H\bigr)\dot\phi
+\bigl(-6H^2-4\dot H+2\frac{k}{a^2}\bigr)\dot\phi^{\,2},
\end{align}
and
\begin{eqnarray}
B(t)&=&
4\dot{\phi}^{4}
- 4\,\dot{\phi}^{2}\ddot{\phi}
+ 8H\dot{\phi}\ddot{\phi}
-12H\dot{\phi}^3
+ H^2\left(-4H\dot\phi+12\dot\phi^2-4\ddot\phi \right)\nonumber\\
&&-\dot{H}\left(-8H\dot \phi+4\dot\phi^2\right)+\frac{k}{a^2}\left(8\dot\phi^2
-12H\dot \phi -4\ddot \phi \right)\nonumber\\
&=&4\dot\phi^{\,4}-12H\dot\phi^{\,3}
+\bigl(8H\dot\phi-4\dot\phi^{\,2}-4H^2-4\frac{k}{a^2}\bigr)\ddot\phi+\bigl(8H\dot H-4H^3-12H\frac{k}{a^2}\bigr)\dot\phi\nonumber\\
&& +\bigl(12H^2-4\dot H+8\frac{k}{a^2}\bigr)\dot\phi^{\,2}.
\end{eqnarray}
Here, one notes that $\mathcal{E}_{\mu\nu}=-\hat{\alpha}\hat{H}_{\mu\nu}$
in comparison to \eqref{fess}. Thus, the field equations \eqref{fescalar} of this theory will be
\begin{eqnarray}\label{fess}
\rho&=&3\rho_1 +\hat{\alpha}\left(A-B\right)=3\rho_{1} +\hat{\alpha}\left[
-3\dot{\phi}^{4}+12H\dot{\phi}^3
-6\left(3H^2 +\frac{k}{a^2}\right)\dot{\phi}^2+12\left( H^3 +H\frac{k}{a^2}\right)\dot\phi
\right],\nonumber\\
p&=&-3\rho_1 +2\rho_2 -\hat{\alpha}A\nonumber\\
&=&-3\rho_1 +2\rho_2 \nonumber\\
&& -\hat{\alpha}\left[\dot\phi^{\,4}
+\bigl(8H\dot\phi-4\dot\phi^{\,2}-4H^2-4\frac{k}{a^2}\bigr)\ddot\phi
+\bigl(8H^3+8H\dot H\bigr)\dot\phi
+\bigl(-6H^2-4\dot H+2\frac{k}{a^2}\bigr)\dot\phi^{\,2} \right].
\end{eqnarray}
One can consider the ordinary matter density $\rho=\sum_i \rho_i,~~i=r,\, m,\, \Lambda$ where $r, m, \Lambda$ stand for   the radiation, matter and cosmological constant, respectively. Hence, defining the dimensionless density parameters
\begin{equation}
\Omega=\frac{\rho}{3H^2},~~~~~~\Omega_k=-\frac{k}{a^2 H^2},~~~~~~ \Omega_\phi:=\frac{\rho_\phi}{3H^2}=-\frac{\hat
\alpha (A-B)}{3H^2},
\end{equation}
the modified Friedmann equation
takes the form
\begin{equation}
\sum_{i=r,\, m,\, \Lambda}\Omega_i+\Omega_k+\Omega_\phi=1.
\end{equation}
The $\hat \alpha$ dependent terms in \eqref{fess} can be interpreted as an effective scalar-sector fluid with density $\rho_\phi=-\hat \alpha (A-B)$ and pressure $p_\phi=\hat\alpha A$, so that the modified Friedmann equation retains the standard sum of $\Omega_{i}$ form. Note that $\Omega_\phi$ is not sign-definite, i.e. depending on the background evolution
$(\dot{\phi},\,\ddot{\phi},\,H,\,\dot{H})$ and the sign of $\hat{\alpha}$, the scalar sector may contribute positively or negatively to the total effective density.

As the result, the ordinary matter and the scalar
field determine the spatial curvature of the universe as follows:
\[
\left\{
\begin{aligned}
\Omega_m + \Omega_r + \Omega_\Lambda +\Omega_\phi < 1 &\iff k=-1, \\
\Omega_m + \Omega_r + \Omega_\Lambda +\Omega_\phi = 1 &\iff k=0, \\
\Omega_m + \Omega_r + \Omega_\Lambda +\Omega_\phi > 1 &\iff k=1.
\end{aligned}
\right.
\]
In this model, constraints on spatial curvature from background observables effectively constrain the combination
$\Omega_k+\Omega_\phi$; therefore, a nonzero $\Omega_\phi$ can mimic curvature in standard analyses.
Finally, one can define an effective scalar equation of state as
$w_\phi=p_\phi/\rho_\phi=-A/(A-B)$, which is generally time-dependent and may cross the acceleration threshold
$w_\phi=-1/3$, or even the phantom divide $w_\phi=-1$, depending on the background evolution.

On the other hand, Eq. (\ref{scalar}) governing the scalar field takes the
form
\begin{eqnarray}\label{fe}
&&\ddot{\phi}\left(
 -H^{2}
 + 2H\dot{\phi}
 - \dot{\phi}^{2}
 - \frac{k}{a^{2}}
\right)
+ H\dot{\phi}^{3}
+\left( \dot{H}
- 3H^{2}\right)\dot{\phi}^{2}
+ \left(3H^{3}- 2H\dot{H}+\frac{k}{a^{2}}H\right)\dot{\phi}\nonumber\\
&&~~= \left(H^{2}+\frac{k}{a^{2}}\right)\left(H^{2}+\dot{H}\right),
\end{eqnarray}
or equivalently
\begin{equation}
\left[\left(\dot{\phi}-H\right)^{2}+\frac{k}{a^{2}}\right]\ddot{\phi}
=
H\left(\dot{\phi}-H\right)\left[\left(\dot{\phi}-H\right)^{2}+\frac{k}{a^{2}}\right]
+\dot{H}\left[\left(\dot{\phi}-H\right)^{2}-\frac{k}{a^{2}}-2H^{2}\right].
\end{equation}
For the spatially flat universe, i.e. $k=0$, the above equation can be simplified by letting $\dot{\phi}=H+\psi^{\frac{1}{3}} $ as
\begin{equation}
\dot{\psi}-3 H \psi=-6H^2 \dot{H},
\end{equation}
which has the general solution for $\psi$ as
\begin{equation}
\psi=-6 e^{3 {\int H dt}}\, \int\, \left(\dot{H}\, H^2\, e^{-3 \int H dt} \right) dt+C\,e^{3 \int H dt},
\end{equation}
where $C$ is any constant.

We then find that, for the case $C=0$, $\phi(t)$
reads as
\begin{equation}
\phi(t)= C_{1}+\ln a-6^{\frac{1}{3}}\, \int e^{\int H dt} \left(\int \left(H^2 \dot{H}\,e^{-3\int H dt}\right)dt \right)^{\frac{1}{3}}dt,
\end{equation}
where $C_{1}$ is an arbitrary constant.

Also, when $H= constant=\Lambda$,  setting  $C \ne 0$, we find
\begin{equation}
\psi=C e^{3 \Lambda t},
\end{equation}
and hence
\begin{equation}
\phi(t)=C_{1}+\Lambda t+C_2 e^{\Lambda t},
\end{equation}
where $C$, $C_1$ and $C_{2}=C^\frac{1}{3}/\Lambda$ are arbitrary integration constants.

One can  consider the following two special cases:
\begin{enumerate}\item  \textbf{An spatially flat universe with a de Sitter type  scale factor ($k=0, ~~a(t)=a_0\exp(\Lambda t)$):}

Here, the solution for $\phi(t)$ is of the form

\begin{equation}
\phi(t)=C_{1}+\Lambda t+C_{2} e^{\Lambda t},
\end{equation}
where $C_1$ and $C_{2}$ are arbitrary integration constants.  The corresponding
energy
density and pressure  read as follows
\begin{eqnarray}
\rho(t)&=&3\Lambda^2+3\hat{\alpha}\Lambda^4\left(1-C_2^{\,4}e^{4\Lambda t}\right),
\nonumber\\
p(t)&=&-3\Lambda^2-\hat{\alpha}\Lambda^4\left(3+C_2^{\,4}e^{4\Lambda t}\right).
\end{eqnarray}

\item\textbf{ An spatially flat universe with a power law scale factor ($k=0,~~ a(t)=a_0 t^n$):}

For $n \ne- 1$, we find

\begin{equation}
\phi=~C_1+n\left[1-\left( \frac{2}{n+1} \right)^{1/3}\right] \ln t.
\end{equation}
Hence, the density and pressure profiles read as
\begin{eqnarray}
\rho(t)&=&
\frac{3n^{2}}{t^{2}} +
\frac{3\hat{\alpha}\, n^{4}}{t^{4}}
\left[1-\left(\frac{2}{n+1}\right)^{4/3}
\right],\nonumber\\
p(t)&=&
\frac{n(2-3n)}{t^{2}}
+\frac{\hat{\alpha}}{t^{4}}
\left[
12n^{3}-3n^{4}
-8n\left(-\frac{2n^{3}}{n+1}\right)^{2/3}
-\left(-\frac{2n^{3}}{n+1}\right)^{4/3}
\right].
\end{eqnarray}
Here, one notes that for
$n=1$ the both the brackets in $\rho$ and $p$ vanish, then the scalar field term drops out. Also, in contrast to the early universe, at late times the scalar field piece becomes subdominant.

For $n=-1$, that means a collapsing solution for the scale factor, we have

\begin{equation}
\phi=C_{1}-\ln(t)-\frac{1}{8}
\left(- 6 \ln t\right)^\frac{4}{3},
\end{equation}

Hence the density and pressure profiles read as
\begin{eqnarray}
\rho(t)
&=&\frac{3}{t^{2}}
+\,\frac{3\hat{\alpha}}{t^{4}}
\left(1-\left(-6\ln t \right)^{\frac{4}{3}}\right),\nonumber\\
p(t)&=& -\frac{5}{t^{2}}
-\frac{\hat\alpha}{t^{4}}\left( \left(-6\ln t\right)^{\frac{4}{3}}-8\left(-6\ln t\right)^{\frac{2}{3}}+15 \right).
\end{eqnarray}

\end{enumerate}

%
  \subsection{Regularized Vector-Tensor 4D Einstein-Gauss-Bonnet Theory}

A new vector--tensor regularization of 4D Einstein-Gauss-Bonnet theory was also proposed in \cite{Fernandes2025}, where the regulating field is a vector
field. This is qualitatively different from the scalar--tensor constructions in the previous subsection and is
motivated by the observation that known scalar--tensor regularizations admit a formulation with improved conformal
properties for the extra field. Ref.~\cite{Fernandes2025} shows that an analogous improved conformal structure
for a vector field arises naturally within Weyl geometry, and uses it to implement a dimensional-regularization limit
directly in four dimensions.

The basic ingredient in the new construction is the local Weyl-gauge transformation
\begin{equation}
g_{\mu\nu}\ \to e^{2\sigma(x)}g_{\mu\nu},
\qquad
W_\mu\ \to W_\mu-\partial_\mu\sigma,
\label{rescaling}
\end{equation}
where $W_\mu$ is the Weyl gauge field. In Weyl geometry, the spacetime connection is taken to be torsionless but
non-metric, obeying
\begin{equation}
\tilde{\nabla}_\lambda g_{\mu\nu}=-2W_\lambda g_{\mu\nu}.
\label{eq:weyl_nonmetricity}
\end{equation}
Equivalently, the Weyl connection $\tilde{\Gamma}^\lambda{}_{\mu\nu}$ differs from the Levi--Civita connection
$\Gamma^\lambda{}_{\mu\nu}$ by
\begin{equation}
\tilde{\Gamma}^\lambda{}_{\mu\nu}
=
\Gamma^\lambda{}_{\mu\nu}
+\delta^\lambda_\mu W_\nu+\delta^\lambda_\nu W_\mu
-g_{\mu\nu}W^\lambda.
\label{eq:weyl_connection}
\end{equation}
A key property is that the combined transformation \eqref{rescaling} leaves $\tilde{\Gamma}^\lambda{}_{\mu\nu}$
invariant; consequently, curvature tensors constructed from $\tilde{\Gamma}^\lambda{}_{\mu\nu}$ enjoy simple conformal
behavior. This provides the vector counterpart of the improved conformal structure encountered in the scalar case.   The reference \cite{Jimenez} provides a comprehensive treatment of how to formulate the counterpart of the GB invariant within the framework of a Weyl connection. In \cite{tanhayi}, the authors demonstrate that, at the quadratic level, the only consistent Weyl-invariant gravitational theory is given by the Weyl-geometric generalization of the Einstein-Gauss-Bonnet theory.

\medskip

Starting from the Gauss--Bonnet invariant in $D$ dimensions,
\begin{equation}
\mathcal{G}
=
R_{\mu\nu\rho\sigma}R^{\mu\nu\rho\sigma}
-4R_{\mu\nu}R^{\mu\nu}
+R^2,
\label{eq:GB_def_vec}
\end{equation}
Ref.~\cite{Fernandes2025} considers in addition the Gauss--Bonnet combination $\tilde{\mathcal{G}}$ built from
the Weyl-connection curvatures. The central structural identity is that $\tilde{\mathcal{G}}$ can be rewritten as
\begin{equation}\label{gb}
\tilde{\mathcal{G}}
=
\mathcal{G}
+(D-3)\nabla_\mu J^\mu
+(D-3)(D-4)\,L,
\end{equation}
where
\begin{align}
J^\mu &=
8G^{\mu\nu}W_\nu
+4(D-2)\Big[
W^\mu\big(W^2+\nabla_\nu W^\nu\big)-W^\nu\nabla_\nu W^\mu
\Big],
\label{eq:J_current}\\[3pt]
L &=
4G^{\mu\nu}W_\mu W_\nu
+(D-2)\Big[
4W^2\nabla_\mu W^\mu+(D-1)W^4
\Big].
\label{eq:L_density}
\end{align}
Here,  $W^2\equiv W_\mu W^\mu$ and $W^4\equiv (W^2)^2$.
This decomposition is the analogue of the scalar conformal expansion used in the previous subsection: the second term in \eqref{gb}
is a total derivative, hence relevant only through boundary data, while the last term carries an explicit $(D-4)$
factor which enables a finite $D\to4$ limit after the usual dimensional-regularization rescaling.

\medskip

In particular, Ref.~\cite{Fernandes2025} defines the finite four-dimensional vector--tensor Gauss--Bonnet
density by
\begin{equation}
\mathcal{L}^{\rm VT}_{\mathcal{G}}
\equiv
\lim_{d\to 4}\frac{\mathcal{G}-\tilde{\mathcal{G}}}{d-4}.
\label{eq:VT_def_limit}
\end{equation}
Using \eqref{gb} and discarding the total derivative, the four-dimensional
limit yields
\begin{equation}
\mathcal{L}^{\rm VT}_{\mathcal{G}}
=
4G_{\mu\nu}W^\mu W^\nu
+8W^2\nabla_\mu W^\mu
+6W^4.
\label{eq:VT_GB_density_4D}
\end{equation}
Hence, the corresponding regularized four-dimensional vector--tensor EGB model is then taken to be
\begin{equation}
S[g,W]
=
\int d^4x\,\sqrt{-g}\,
\Big(
R-\alpha\,\mathcal{L}^{\rm VT}_{\mathcal{G}}
\Big)+ S_{m},
\label{eq:VT_action}
\end{equation}
with coupling $\alpha$ (of dimension length$^2$). This theory
belongs to the generalized Proca class and therefore propagates the appropriate number of degrees of freedom with
equations of motion containing no derivatives higher than second order \cite{heisenberg2014,
heisenberg2017}. For the most general covariant ghost-free scalar-vector-tensor (SVT) theories with second-order equations of motion, encompassing both the $U(1)$ gauge-invariant and gauge-broken cases, and reducing to generalized Proca and Horndeski interactions in appropriate limits, see also \cite{heis3}.

\medskip

The field equations of this theory can be obtained by the variation with
respect to the metric and vector field, respectively, as
 \begin{eqnarray}\label{fes}
G_{\mu\nu} &=& -\,\alpha\, H_{\mu\nu} +T_{\mu\nu}, \nonumber\\
G_{\mu}{}^{\rho} W_{\rho} + 3 W^{2} W_{\mu} + 2 W_{\mu} (\nabla_{\rho} W^{\rho})
 - 2 W^{\rho} (\nabla_{\mu} W_{\rho}) &=& 0,
 \end{eqnarray}
where  $H_{\mu\nu}$ tensor is
\begin{eqnarray}
H_{\mu\nu} &=& 2 R_{\mu\nu} W^{2}
 - 2 R_{\nu\rho} W^{\rho} W_{\mu}
 - 2 R_{\mu\rho} W^{\rho} W_{\nu}
 - 4 R_{\mu\rho\nu\sigma} W^{\rho} W^{\sigma}
 + 2 R W_{\mu} W_{\nu}
\nonumber\\
&&\quad - 12 W^{2} W_{\mu} W_{\nu}
 - 8 W_{\mu} W_{\nu} \nabla_{\rho} W^{\rho}
 - 2 (\Box W_{\mu}) W_{\nu}
 - 2 W_{\mu} \Box W_{\nu}
 - 4 \nabla_{\rho} W_{\mu} \nabla^{\rho} W_{\nu}
\nonumber\\
&&\quad + 8 (\nabla_{\mu} W_{\rho}) W_{\nu} W^{\rho}
 + 2 (\nabla_{\mu} W_{\rho})(\nabla^{\rho} W_{\nu})
 + 2 (\nabla_{\mu} W_{\nu})(\nabla_{\rho} W^{\rho})
 + 2 (\nabla_{\mu}\nabla_{\rho} W^{\rho}) W_{\nu}
\nonumber\\
&&\quad + 2 W^{\rho} (\nabla_{\mu}\nabla_{\rho} W_{\nu})
 + 8 W^{\rho} W_{\mu} (\nabla_{\nu} W_{\rho})
 + 2 (\nabla_{\rho} W_{\mu})(\nabla_{\nu} W^{\rho})
 - 4 (\nabla_{\mu} W_{\rho})(\nabla_{\nu} W^{\rho})
\nonumber\\
&&\quad + 2 (\nabla_{\nu} W_{\mu})(\nabla_{\rho} W^{\rho})
 + 2 W_{\mu} (\nabla_{\nu}\nabla_{\rho} W^{\rho})
 + 2 W^{\rho} (\nabla_{\nu}\nabla_{\rho} W_{\mu})
 - 4 W^{\rho} (\nabla_{\mu}\nabla_{\nu} W_{\rho})
\nonumber\\
&&\quad + g_{\mu\nu}\Big(
 4 R_{\rho\sigma} W^{\rho} W^{\sigma}
 - R W^{2}
 + 3 W^{4}
 - 8 W^{\rho} W^{\sigma} (\nabla_{\sigma} W_{\rho})
 - 2 (\nabla_{\rho} W^{\rho})^{2}
\nonumber\\
&&\qquad\qquad
 - 4 W^{\rho} ( \nabla_{\sigma}\nabla_{\rho} W^{\sigma} - \Box W_{\rho})
 + 2 \nabla_{\sigma} W_{\rho} \big( 2 \nabla^{\sigma} W^{\rho} - \nabla^{\rho} W^{\sigma} \big)
\Big),
\end{eqnarray}
and  $T_{\mu\nu}=pg_{\mu\nu} +\left( \rho +p\right)u_\mu u_\nu$ is the ordinary perfect fluid energy-momentum tensor from $S_m$.

Imposing the homogeneous and isotropic ansatz on the Proca field
\begin{equation}
W_{\mu}=w(t)\,u_{\mu}, \qquad u_{\mu}u^{\mu}=-1,
\end{equation}
the 2nd equation in (\ref{fes}) in 4D reduces to the following algebraic
equation for $w(t)$
 \begin{equation}
   w^2-2wH+\rho_1=0,
  \end{equation}
 leading to
 \begin{equation}\label{wsol}
 w(t)=H\pm\frac{\sqrt{-k}}{a}
 \end{equation}

Hence, $w(t)$ can be determined in terms of the scale factor $a(t)$ and the
spatial curvature $k$. One also notes that the spatial curvature here cannot be positive for a real valued scalar $w(t)$ function.

In order to verify the validity of our theorem in this theory in 4-dimensions,  one can use
\[
W^2=-w^2,\qquad
\nabla_\mu W_\nu=\dot w\,u_\mu u_\nu+Hw\,(g_{\mu\nu}+u_\mu u_\nu),
\qquad
\nabla_\mu W^\mu=-\dot w+3Hw,
\]
together with the wave operator
\[
\Box W_\mu
=\left(-\ddot w+3H\dot w+3H^2 w\right)\,u_\mu,
\]
and curvature contractions
\begin{eqnarray}
R_{\mu\rho\nu\sigma}W^\rho W^\sigma
&=& w^2\,R_{\mu\rho\nu\sigma}u^\rho u^\sigma
= w^2\,\left(\rho_2-\rho_1\right)\,(g_{\mu\nu}+u_\mu u_\nu)\nonumber\\
R_{\mu\nu}W^\mu W^\nu
&=&w^2\,R_{\mu\nu}u^\mu u^\nu
= 3w^2\,\rho_2-\rho_1\big),
\nonumber\\
R_{\mu\nu}W^2
&=& -w^2 R_{\mu\nu}
= -w^2\Big[g_{\mu\nu}\big(3\rho_1-\rho_2\big)
+2u_\mu u_\nu \rho_2\Big],
\nonumber\\
R_{\mu\rho}W^\rho W_\nu
&=& 3w^2 \big(\rho_1-\rho_2\big)\,u_\mu u_\nu,
\nonumber\\
R\,W_\mu W_\nu
&=& R\,w^2\,u_\mu u_\nu
=3\big(4\rho_1-2\rho_2\big)w^2\,u_\mu u_\nu,
\nonumber\\
R\,W^2
&=& -w^2 R
= -3\big(\rho_1-2\rho_2\big)w^2.
\end{eqnarray}
These identities ensure that every term in the tensor
$H_{\mu\nu}$ collapses to the basis
$\{g_{\mu\nu}, u_\mu u_\nu\}$, in accordance with our closure theorem.
The explicit form of $H_{\mu\nu}$ will be
\begin{equation}
H_{\mu\nu}=A(t)g_{\mu\nu} + B(t)u_\mu u_\nu,
\end{equation}
where
\begin{align}
A(t)&= 2\ \left(-2\dot H - 3H^2+\frac{k}{a^2}
-4\dot w \right)w^2+8H\dot w w +3w^4,\nonumber\\
B(t)&= 4\left(  5\dot H + 3H^2 + \frac{2k}{a^2}-2\dot
w\right)w^{2}+8H\dot w w -24H w^3 +12 w^{4}.
\end{align}
Hence the field equations of the theory read as
\begin{eqnarray}\label{vfes}
\rho&=&3\rho_1 +\alpha( B-A)=3\rho_1+\alpha\left[
6\ \left(4\dot H + 3H^2+\frac{k}{a^2} \right)w^2-24Hw^{3} +9w^4 \right],\nonumber\\
p&=&-3\rho_1+2\rho_2+\alpha A=-3\rho_1+2\rho_2+
\alpha\left[2\ \left(-2\dot H - 3H^2+\frac{k}{a^2}
-4\dot w \right)w^2+8H\dot w w +3w^4  \right].
\end{eqnarray}
Similar to the scalar field case, here we can also decompose the ordinary
energy density to the radiation, matter, and cosmological constant contributions, while the
$\alpha$--sector as an vector field fluid with
$\rho_w=-\alpha(B-A),\,\, p_w=-\alpha A.$
Defining the dimensionless density parameters
\begin{equation}
\Omega_i=\frac{\rho_i}{3H^2},\qquad
\Omega_k=-\frac{k}{a^2H^2},\qquad
\Omega_w=\frac{\rho_\alpha}{3H^2},
\end{equation}
the modified Friedmann equation can be written in the standard closure form
\begin{equation}
\sum_{i=r,\,m,\,\Lambda}\Omega_i+\Omega_k+\Omega_w=1.
\end{equation}
Since $\Omega_w$ is not sign--definite and generally evolves in time,
constraints on the spatial curvature inferred from background observables
effectively constrain the combination $\Omega_k+\Omega_w$. As a result,
a nonvanishing $\Omega_w$ may mimic spatial curvature in analyses that
assume a purely standard cosmological sector. The spatial geometry is thus
determined by
\[
\left\{
\begin{aligned}
\Omega_m+\Omega_r+\Omega_\Lambda+\Omega_w<1 &\iff k=-1,\\
\Omega_m+\Omega_r+\Omega_\Lambda+\Omega_w=1 &\iff k=0,\\
\Omega_m+\Omega_r+\Omega_\Lambda+\Omega_w>1 &\iff k=+1.
\end{aligned}
\right.
\]

For an spatially flat universe,  using  \eqref{wsol}, the field equations \eqref{vfes} simplify
to
\begin{eqnarray}
\rho&=&3 H^2+3 \alpha H^2\left( 8 \dot H +  H^2\right),\nonumber\\
p&=&-3H^2 -2\dot H -\alpha H^2\left(4\dot H + 3H^2 \right).
\end{eqnarray}
Hence, the following two cases can be interesting.

\begin{enumerate}
\item \textbf{An spatially flat universe with a de Sitter type  scale factor
($k=0,~a(t)=a_0\exp(\Lambda t)$):}

Here the energy density and pressure of the vector field read as
\begin{equation}
\rho=3 \Lambda^2(1+\alpha \Lambda^2),~~~~p=-3 \Lambda^2(1+\alpha \Lambda^2).
\end{equation}
Here, one finds the simple equation of state parameter $w=\frac{p}{\rho}=-1.$
\item  \textbf{An spatially flat universe with a power law scale factor ($k=0, ~~a(t)=a_0t^{n}$):}

Here, the energy density and pressure  take the form
\begin{eqnarray}
\rho(t)=\frac{3n^{2}}{t^{2}}+\alpha\,\frac{3n^{3}(n-8)}{t^{4}},~~~~p(t)=\frac{2n-3n^{2}}{t^{2}}+\alpha\,\frac{n^{3}(4-3n)}{t^{4}}
\end{eqnarray}
In this case the equation of state parameter is complicated.

\end{enumerate}
\section{Conclusion}

In this work, we have extended our previous theorem for FLRW spacetimes in generic metric gravity theories based on Riemannian geometry to a broader class of modified gravitational models including scalar and vector degrees of freedom. More specifically, we considered scalar--vector--tensor theories whose Lagrangians may depend on the metric, the curvature tensor and its covariant derivatives, as well as scalar and vector fields and their derivatives
at any order. We proved that, under the symmetry conditions imposed by FLRW geometry, the metric field equations necessarily reduce to the Einstein form with an effective perfect-fluid source.

\vspace{0.5cm}
\noindent
This establishes that FLRW metrics retain a universality property even in the presence of additional scalar and vector fields. The essential point
proved here is that the tensorial structure of the metric field equations is fixed entirely by the symmetry of the spacetime and is therefore independent of the detailed form of the gravitational Lagrangian. At the same time, the effective energy density and pressure remain theory dependent, and hence so does the resulting cosmological evolution.

\vspace{0.5cm}
\noindent
To illustrate the general theorem proved in the present work, we examined recently proposed scalar--tensor and vector--tensor theories arising from regularized four-dimensional Einstein--Gauss--Bonnet gravity. In both cases, the modified field equations in an FLRW background reduce to the perfect-fluid form predicted by the theorem, with the effective energy density and pressure determined by the dynamics of the corresponding scalar or vector fields. These examples show explicitly that even theories with higher-curvature origins and additional dynamical fields obey the same structural constraints imposed by cosmological symmetry.

\vspace{0.5cm}
\noindent
This research does not propose a new parametrization of already known equations. Rather, it establishes a \emph{theory-independent structural theorem} for a very broad class of scalar--vector--tensor theories. The theorem states that, under the symmetry assumptions of FLRW spacetime and the corresponding restrictions on the scalar and vector fields, the metric field equations must necessarily take the form of Einstein equations sourced by an effective perfect fluid. This is not assumed and not obtained by a case-by-case rearrangement
or any manipulations; it is proved directly from the tensorial structure permitted by FLRW symmetry.

\vspace{0.5cm}
\noindent
Theorem 1 and 2 places FLRW metric within the broader framework of \emph{universal metrics}. As discussed earlier, universality is known for certain special geometries such as gravitational plane waves, pp-wave metrics, and Kerr--Schild--Kundt metrics, where arbitrary higher-curvature corrections collapse to a much simpler form. Our earlier work showed that FLRW metrics also have such a universality property for generic metric theories constructed from curvature and its covariant derivatives. The present paper extends this statement to scalar--vector--tensor theories that is mathematically nontrivial and conceptually important, because it demonstrates that universality persists even after enlarging the field content significantly.

\vspace{0.5cm}
\noindent
The theorem 1 and 2  here has an important physical value as it identifies clearly what is universal and what is not. The universal part is the perfect-fluid form of the metric field equations at the homogeneous and isotropic level. The non-universal part is encoded in the theory-dependent effective density and pressure, which determines the actual evolution of the scale factor. In this way, the theorem separates model-independent background structure from model-dependent dynamics. We believe this is useful both conceptually and practically.

\vspace{0.5cm}
\noindent
Our result therefore provides a general explanation for why a wide variety of modified gravity theories share the same perfect-fluid structure at the level of homogeneous and isotropic cosmology. While the detailed dynamics remains model dependent, the theorem shows that the background tensorial form of the field equations is universal. This universality property offers a useful framework for analyzing the background behavior of broad classes of modified gravity theories and suggests that genuinely distinguishing features are expected to arise beyond the FLRW background, for example in cosmological perturbations, anisotropic spacetimes, or gravitational-wave propagation.

\section{Acknowledgment}
The work of YH was supported by the Science Academy's Young Scientist Awards Program (BAGEP) 2025.


\end{document}